\documentclass[reprint,superscriptaddress,nofootinbib,amsmath,amssymb,aps,prd]{revtex4-2}
\usepackage{bm, graphics, graphicx, epsfig, soul, latexsym, hyperref ,multirow}
\usepackage{comment}
\usepackage{subfigure}
\usepackage{graphicx,color}
\usepackage{dcolumn}
\usepackage{bm,amsmath, amssymb,}
\usepackage{mathrsfs}
\usepackage{bigints}
\usepackage{float}
\usepackage{multirow}
\usepackage{placeins}
\usepackage{subfigure}
\usepackage[normalem]{ulem}
\usepackage{booktabs}
\usepackage{tabularx}

\usepackage{amsmath}

\newcommand{\be}{\begin{equation}}
\newcommand{\ee}{\end{equation}}

\begin{document}
%TC:ignore
\title{Spin-induced quadrupole moment based test for eccentric binaries}
\author{N. V. Krishnendu}
\email{k.naderivarium@bham.ac.uk}
\affiliation{School of Physics and Astronomy, University of Birmingham, Edgbaston, Birmingham, B15 2TT, UK}
\date{\today}
\begin{abstract}
%%%%%
 The spin-induced quadrupole moment-based test of black hole nature is routinely used to probe the true nature of detected binary signals, assuming a circular orbit. We extend the applicability of the method to binaries in eccentric orbits. Considering simulated signals of varying masses, spins, and signal strengths, we demonstrate how the systematic errors resulting from neglecting orbital eccentricity compare with the statistical errors, using a semi-analytic Fisher matrix-based formalism that accounts for both current and future detectors. Further, we quantify the systematic errors by developing a Bayesian inference framework for the current detector network. The inspiral-only aligned spin gravitational wave waveform model for eccentric binaries, \texttt{TaylorF2Ecc}, is employed. For the current detector network, neglecting an initial eccentricity of $e_0^{\rm inj}=0.1$ defined at $20\,\mathrm {Hz} $ can lead to a serious bias in binary parameter inference. Notably, a nearly equal-mass, moderately spinning binary black hole in an eccentric orbit can be identified as a non-black hole binary with extreme spins and asymmetric masses. We demonstrate the criticality of biased estimates that may arise when neglecting the orbital eccentricity while performing tests of black hole nature and discuss prospects. 
%%%%%
\end{abstract}
%%%%%
\maketitle
%TC:endignore
%%%%%%%%%%%%%%%%%%%%%%%%%%%%%%%%%%%%%%%%%%%%%%%%%%%%%%%%%%%%%%%%%%%%%%%%%%%%%%
%%%%%%%%%%%%%%%%%%%%%%%%%%%%%%%%%%%%%%%%%%%%%%%%%%%%%%%%%%%%%%%%%%%%%%%%%%%%%%
%%%%%%%%%%%%%%%%%%%%%%%%%%%%%%%%%%%%%%%%%%%%%%%%%%%%%%%%%%%%%%%%%%%%%%%%%%%%%%
\section{Introduction}
\label{intro}
Gravitational wave (GW) measurements of spin induced multipole moments provide an unique way to perform tests of nature of compact binaries~\cite{Krishnendu:2017shb, Krishnendu:2018nqa, Krishnendu:2019tjp}, hence to distinguish binary black hole (BH) mergers from merger of non-BH compact objects, including neutron stars and other exotic stars~\cite{Ryan97, Cardoso:2017cfl, CardosoTidal2017, Pacilio:2020jza, Uchikata2015, Uchikata2016, Johnson-Mcdaniel:2018cdu}. The method relies on using GW measurements of spin-induced multipole moments to distinguish between different classes of compact objects, as the values of spin-induced multipole moments are only a function of mass and spin for Kerr BHs, and hence the coefficients are always unity. In contrast, for non-BH compact objects and BH solutions in alternative theories of gravity, the spin-induced multipole moments are also functions of the specific equation of state (internal structure) along with the mass and spin. Hence, the spin-induced multipole moment coefficients will not be the same and will not be equal to one. For example, numerical simulations of slowly spinning neutron stars show that the spin-induced quadrupole moment values range between $~\sim 2-14$ depending upon the mass and spin choices of the configuration~\cite{PappasMultipole2012, Pappas:2012qg, Yagi:2014bxa, Sotiriou:2004bm, Apostolatos:2006vu}. Similarly, for rotating self-interacting boson stars, the spin-induced quadrupole moments can be as large as $\sim 150$, and the spin-induced octupole moments can take even larger values $\sim 200$~\cite{Ryan97, Pacilio:2020jza}. 

The detected binary signals are analysed by performing the spin-induced quadrupole moment-based test of the nature of compact objects, both across the spinning binaries in the catalogue and for specific exceptional events. The most stringent constraints are obtained from GW241011~\cite{LIGOScientific:2025brd, Krishnendu:2025rud}, mainly due to its loud inspiral, a rapidly spinning primary, and the unequal mass ratio. The constraints derived from real data depend strongly on the underlying binary parameters. As an example, the spin-induced quadrupole moment estimates for the asymmetric systems GW190814~\cite{GW190814} and GW190412~\cite{GW190412, GWTC-2-TGR} were used to put constraints on the nature; yet only GW190412 yields informative bounds due to its pronounced spin effects, whereas GW190814 remains effectively uninformative owing to its slow-spinning components. 

The test is performed by introducing phenomenological deviation coefficients in the inspiral regime of the binary dynamics~\cite{Krishnendu:2019tjp, Saini:2023gaw}, representing the spin-induced multipole moment parameters. For a spinning compact binary system, post-Newtonian approximation (PN) provides an accurate description of the dynamics in the frequency domain~\cite{Marsat2015, M3B2013, MKAF16}. Specifically, for the spin-induced quadrupole moment parameters, the coefficients start to appear at the second post-Newtonian order (2 PN) along with the self-spinning (quadratic spin terms). The initial demonstration of the test focused on post-Newtonian inspiral-only aligned-spin waveform models, where the dynamics are accurate up to a frequency which is equal to the last stable circular orbit frequency of a Kerr BH and the spin orientations of the binary are restricted to aligned/anti-aligned with the orbital angular momentum axis of the binary~\cite{Krishnendu:2017shb}. Later, this method was applied on phenomenological waveform models with spin-precession effects~\cite{Krishnendu:2019tjp, GWTC-2-TGR, GWTC-3-TGR}, where the spin vectors are allowed to orient freely, and the model is valid in the post-inspiral regime as well. The applicability of the method is restricted to inspiral-dominated signals, as the validity of the PN approximation does not extend to the post-inspiral part of the dynamics. We lack a more accurate waveform model that accounts for spin-induced multipole moment coefficients in the merger and ringdown regimes. The numerical simulations of spinning BH binaries account for spin-induced effects; however, a phenomenological way of constructing an inspiral-merger-ringdown waveform model with parametrised coefficients in the merger-ringdown is not yet available. The latest waveform models employed for the spin-induced quadrupole moments use the state-of-the art phenomenological waveform models with full spin-precession effects (two-spin precession) and includes higher order modes $(\ell, |m|) = (3, 3),(4, 4),(2, 1),(3, 2) $  along with $(\ell, |m|) = (2, 2)$ the dominant mode~\cite{Divyajyoti:2023izl} as the GW catalogue contains a variety of binary sources, including the ones from low mass gap~\cite{GW230529, LIGOScientific:2020aai, GW190814, Schmidt:2024hac}, spin-precession effects~\cite{NS_BH_GW200105_GW200115,gwtc-3-catalogue, GWTC-2.1-catalog, GWTC-2-catalog, GWTC-4-catalogue} and orbital eccentric effects~\cite{Morras:2025xfu}, and many more such events to come~\cite{GWTC-4-pop, LIGOScientific:2025yae, LIGOScientific:2025hdt, LIGOScientific:2025snk}, studying the effect of orbital eccentricity in performing the spin-induced quadrupole moment coefficients based test of BH nature is crucial.

\subsection{GWs from eccentric binaries}
A compact binary system undergoes three major phases during its evolution: inspiral, merger and ringdown. During the inspiral and ringdown phases, the emitted GW signal is modelled by perturbative techniques based on PN theory and BH perturbation theory, respectively. While modelling the merger, the most energetic and relativistic regime requires numerical relativity techniques. The inspiralling orbits can be eccentric depending on the distinct astrophysical environment in which the binary resides. However, binaries in eccentric orbits circularise over time. Hence, in the last few cycles close to the merger, the binary is usually in circular orbit~\cite{PM63, Pe64}. Measuring orbital eccentricity, along with other binary intrinsic and extrinsic parameters, is an important differentiator for the various binary formation channels. For example, binaries formed in dense stellar environments or through Kozai-Lidov processes will be in eccentric orbit $({e}_0 \geq ~\sim 0.1)$ at frequencies $~\sim 10 \rm{Hz}$~\cite{Samsing:2017xmd}. Hence we can detect these eccentric systems using the current detectors~\cite{OLeary:2005vqo, Fragione:2019hqt, Kumamoto:2018gdg, Gondan:2020svr, Gondan:2017hbp,Hoang:2017fvh,VanLandingham:2016ccd,Osburn:2015duj,2012Natur.490...71S, Chomiuk:2013qya, Antonini:2015zsa,Rodriguez:2016kxx,LIDOV1962719,1962AJ.....67..591K, Zwick:2025ine}.

As the GW transient catalogue already contains more than two hundred events, excluding the public alerts of more than three hundred events from the fourth observing run alone, we have evidence for spinning, spin-precessing, mass asymmetric, eccentric binaries~\cite{Morras:2025xfu, Kacanja:2025kpr, Romero-Shaw:2025vbc, Planas:2025plq, Planas:2025jny, Dhurkunde:2023qoe}. Eccentricity and spin-induced orbital precession are the two complex features in binary dynamics. Employing a waveform model including both these effects is a must for correctly identifying binary signals from noisy data and estimating the astrophysical and fundamental physics properties of observed binaries~\cite{Ramos-Buades:2023yhy, Iglesias:2022xfc, Gupte:2024jfe, OShea:2021faf, Romero-Shaw:2022xko, Romero-Shaw:2021ual}. Excluding orbital eccentricity effects while generating templates for GW detection and parameter inference can lead to missing signals and mis-identifying source properties, affecting sources with large eccentricity~\cite{Phukon:2024amh, Favata:2021vhw, Favata2013}. 

 These modelling systematics have direct implications for tests of general relativity (GR), as demonstrated in several recent investigations. The potential for binary BH eccentricity to mimic apparent violations of GR in the routinely employed tests has been investigated in~\cite{Narayan:2023vhm} and reported that signals from eccentric binary BHs with eccentricities as low as 0.1 at $\rm{17Hz}$ can induce deviations in standard tests, including parametrized PN tests, modified dispersion relations, and inspiral–merger–ringdown consistency tests. Notably, even smaller eccentricities, around 0.05 at $\rm{17Hz}$, are found to be sufficient to produce apparent deviations from GR in some cases. A similar study, based on the Fisher information matrix formalism focusing on the parametrized tests of GR, concluded that the systematic errors due to orbital eccentricity can dominate statistical errors on the deviation parameters for eccentricities of $\sim 0.04$ at $\rm{10 Hz}$ in LIGO/Virgo band for binary BHs, and $\sim 0.008$ for binary neutron stars~\cite{Saini:2022igm}. They also reported that with third-generation detector sensitivities, ten times smaller eccentricities can lead to the same bias. The eccentricity-induced biases in the inspiral-merger-ringdown consistency test are investigated in Ref.~\cite{Shaikh:2024wyn} by looking at GW150914-like simulations and finding that the eccentricity of $\sim 0.055$ at $\rm{25 Hz}$ can lead to biased estimates at $68\%$ CI. Another study  focusing on binary neutron star parameter estimation and tidal deformability measurements showed that orbital eccentricity values less than $\sim 10^{-3}$ can cause systematic bias in tidal deformability measurements, leading to incorrect equation of state estimations of neutron stars. In addition, this can bias other source frame parameters of the binary, including the eccentricity parameter, which is a free parameter in the parameter estimation, and can worsen the errors by a factor $\leq 2$~\cite{DuttaRoy:2024aew}. 
 
\subsection{This Work}
In this paper, for the first time, we quantify the implications of excluding orbital eccentricity in performing a spin-induced quadrupole moment test for the nature of BHs. We perform systematic bias studies using the Fisher information matrix and Bayesian inference, employing simulated binary signals. We restrict our analysis to the inspiralling binaries in eccentric orbits with varying masses, spins and signal strengths. The paper is organized as follows. We provide details of the systematic bias studies and their results in Sec.~\ref{sec:systematic_errors}. Following that, we detail the Bayesian inference method and its findings in Sec.~\ref{sec:BayesianResults}. We conclude with a discussion in Sec .~\ref {sec:summary}.

\section{Systematic and statistical errors}
\label{sec:systematic_errors}
\subsection{Recap: spin-induced quadrupole moment based test of BH nature}
The GW measurements of the spin-induced multipole moment parameters can distinguish between BHs and non-BH compact objects~\cite{Krishnendu:2017shb, GWTC-2-TGR, GWTC-3-TGR}. The leading order effect is the spin-induced quadrupole moment. For a compact object with mass $m$ and dimensionless spin $\boldsymbol{\chi}$ the spin-induced quadrupole scalar $Q$ is,
\begin{equation}
    Q=-\kappa\, \boldsymbol{\chi}^3 m^3,
\end{equation}
where the proportionality constant, $\kappa$ is unity for Kerr BHs $\kappa_{\rm BH}=1$. In addition, for non-BH compact stars such as neutron stars and boson stars, the $\kappa$ value is determined by the specific equation states or the internal composition~\cite{Pappas:2012qg, Laar97, Ryan97}. Notably, a binary system will be characterised by the individual spin-induced quadrupole coefficients, $\kappa_1$ and $\kappa_2$. While performing a null test for binary BH nature, it is proposed to measure the symmetric combination of $\kappa_1$ and $\kappa_2$, $\kappa_s=1/2(\kappa_1+\kappa_2)$, keeping the anti-symmetric combination to zero~\cite{Krishnendu:2017shb}. The deviation parameters $\delta\kappa_i$ is related to $\kappa_i$ by,
\begin{equation}
    \kappa_i=1+\delta\kappa_i,
\end{equation}
hence, $\delta\kappa_i=0$ for Kerr BHs. 
Due to the recent advances in analytical modelling of spinning compact objects, the spin-induced quadrupole moment coefficients are available at leading and sub-leading orders in GW waveform models~\cite{Divyajyoti:2023izl, MKAF16, Vines:2016qwa}. The spin-induced multipole moment test for BH nature is employed to assess the properties of detected compact objects~\cite{Saini:2022igm, GWTC-2-TGR, GWTC-3-TGR}, assuming circular orbits. However, its implications for binaries on eccentric orbits remain largely unexplored~\cite{Sridhar:2024zms, Naqvi:2025gly}.
\subsection{\texttt{TaylorF2Ecc} waveform model}
The \texttt{TaylorF2Ecc} waveform is a frequency-domain GW waveform model that extends the quasi-circular {\texttt{TaylorF2}~\cite{Buonanno:2009zt} approximant by including small-eccentricity corrections using the PN expansions. This model includes eccentricity-dependent phase corrections up to 3PN order, making it suitable for modelling compact binary inspirals with low to moderate eccentricities~\cite{Moore:2016qxz}. The model assumes adiabatic inspiral and employs the stationary phase approximation to express the waveform in the Fourier domain~\cite{SPA}. Moreover, the model includes aligned-spin effects of the binary up to $\rm{3.5PN}$ order along with the spin-induced quadrupole moment contributions at $\rm{2PN}$ and $\rm{3PN}$ orders. Hence, \texttt{TaylorF2Ecc} waveform model provides an accurate description for inspiralling aligned-spin compact binary signals in eccentric orbits and has been employed in the literature while performing systematic bias studies and estimating eccentricity parameter along with other binary parameters~\cite{DuttaRoy:2024aew, Saini:2022igm, Phukon:2024amh, Favata08}.

The frequency domain GW waveform is decomposed into an amplitude and a phase,
\begin{equation}
    \tilde{h}(f)=\mathcal{C} A f^{-7/6}e^{\psi(f)}
\end{equation}
$\psi(f)$ is the phase and $A$ is the amplitude of the wave. The amplitude is $A\propto M_c^{5/6}/D_L$, where $M_c$ is the chirp mass related to component masses $m_1$ and $m_2$ as, $M_c={(m_1 m_2)^{3/5}}{(m_1+m_2)^{-1/5}}$ and $D_L$ is the luminosity distance to the source. We restrict the waveform's amplitude at the leading order for the restricted waveform model. The phase, $\psi(f)$, in our case,  is a PN series expansion that accounts for the spin-induced quadrupole moment corrections up to 3PN order and the eccentricity effects up to 3PN order. 

\subsection{Statistical and systematic errors using Fisher information matrix analysis}
The Fisher information matrix formalism is a semi-analytic approximation technique to calculate the statistical and systematic errors on binary parameters at $1-\sigma$ uncertainty, provided the true parameter values. It approximates Bayesian inference-based posterior probability distributions with uniform prior assumptions on the parameters. There have been many attempts to compare the outcomes and limitations~\cite{Kumar:2025nwb, Dupletsa:2024gfl, Porter:2015eha}.
For a binary signal described by the set of parameters $\vec{\theta}$, waveform model $\tilde{h}(f)$ observed using a detector configuration characterized by the noise power spectral density $\rm S_{n}(f)$, the Fisher information matrix is defined as,
\begin{equation}
    \Gamma_{ij}=2 \int_{\rm f_{lower}}^{\rm f_{upper}} \frac{\partial_i \tilde{h}(f)\, \partial_j \tilde{h}^*(f)+\partial_j \tilde{h}(f)\, \partial_i \tilde{h}^*(f)}{\rm S_{n}(f)},
    \label{eq:fisher}
\end{equation}
$\partial_i$ is the partial derivative with respect to the $i^{th}$ parameter, $\vec{\theta_i}$. The lower and upper cut-off of the integral is fixed by the sensitivity of the detector and the signal model. For example, the $\rm f_{lower}$ is $10-20\rm Hz$ for LIGO-type detectors, and $\rm f_{upper}$ is estimated to be the inner-most stable circular frequency of Kerr BHs when analyzing binary BHs with an inspiral-only waveform model. The statistical errors on binary parameters $\vec{\theta}$ is,
\begin{equation}
    \sigma_{\vec{\theta_i}}=\sqrt{C_{ii}}=\sqrt{\Gamma_{ii}^{-1}}.
\end{equation}
The inverse of the Fisher matrix is the covariance matrix, $C_{ij}$~\cite{Vallisneri07}. 

The statistical errors capture the error in estimating the parameter if there are no known systematic errors. That is the signal that provides the best model for the binary dynamics. However, this is no longer the case if the signal is only an {\it approximate} model of the binary dynamics and there is a more accurate {\it true} model. In this case, one must account for the systematic errors arising from the model uncertainties. Additionally, ensure that the systematic errors are well within the statistical uncertainties and do not compromise the findings. Given a true waveform model, 
\begin{equation}
    \tilde{h}^{\rm T}(f) = \mathcal{C} A^{\rm T}\, f^{-7/6}\,e^{\psi^{\rm T}(f)},
\end{equation}
and an approximate model,
 \begin{equation}
    \tilde{h}^{\rm app}(f) = \mathcal{C} A^{\rm app}\,f^{-7/6}\,e^{\psi^{\rm app}(f)},
\end{equation}
the Fisher information matrix also provides a framework for estimating the systematic errors.
For a binary parameter $\vec{\theta_i}$, the systematic error is the difference between the true error $\vec{\theta_i}^{\rm T}$ estimated using $\tilde{h}^{\rm T}(f)$ and the best-fit error $\vec{\theta_i}^{\rm app}$ estimated using $\tilde{h}^{\rm app}(f)$,
\begin{equation}
    \Delta\vec{\theta_i}=\vec{\theta_i}^{\rm T}-\vec{\theta}^{\rm app}.
\end{equation}
Employing the Fisher information matrix formalism, the systematic error $\Delta\vec{\theta_i}$ can be calculated as~\cite{Favata2013},
\begin{equation}
    \Delta\vec{\theta_i}=4\, \left(A^{\rm app}\right)^2\,C_{ij}\int_{\rm f_{lower}}^{\rm f_{upper}} df \frac{f^{-7/3}}{\rm S_{n}(f)}\, \left(\psi^{\rm T}-\psi^{\rm app}\right)\, \partial_j\psi^{\rm app}
\end{equation}
Notice that the covariance matrix $C_{ij}$ is evaluated using $\tilde{h}^{\rm app}(f)$.

To estimate the eccentricity-induced systematic biases in spin-induced quadrupole moment measurements, we use the quasi-circular waveform model as an approximate model and the true model as one that describes a binary in an eccentric orbit. To demonstrate the systematic errors as a function of eccentricity, we consider three detector configurations~\cite{Babak:2021mhe,LIGOScientific:2016wof,Reitze:2019iox,Punturo:2010zza,Hild:2010id,Sathyaprakash:2011bh}: i) LIGO sensitivity with $\rm f_{lower}=20\rm{Hz}$, ii) Cosmic Explorer sensitivity representing next-generation detector sensitivity with $\rm f_{lower}=5\rm{Hz}$ and iii) proposed space-based detector sensitivity corresponding to the LISA configuration with $\rm f_{lower}=10^{-5}\rm{Hz}$. For this study, sensitivity curves detailed in Ref.~\cite{KY19} are used. 
%%%%%%%%%%%%%%%%%%%%%%%%% FIGURE
\begin{figure*}[t]
    %\centering
    \includegraphics[width=\textwidth]{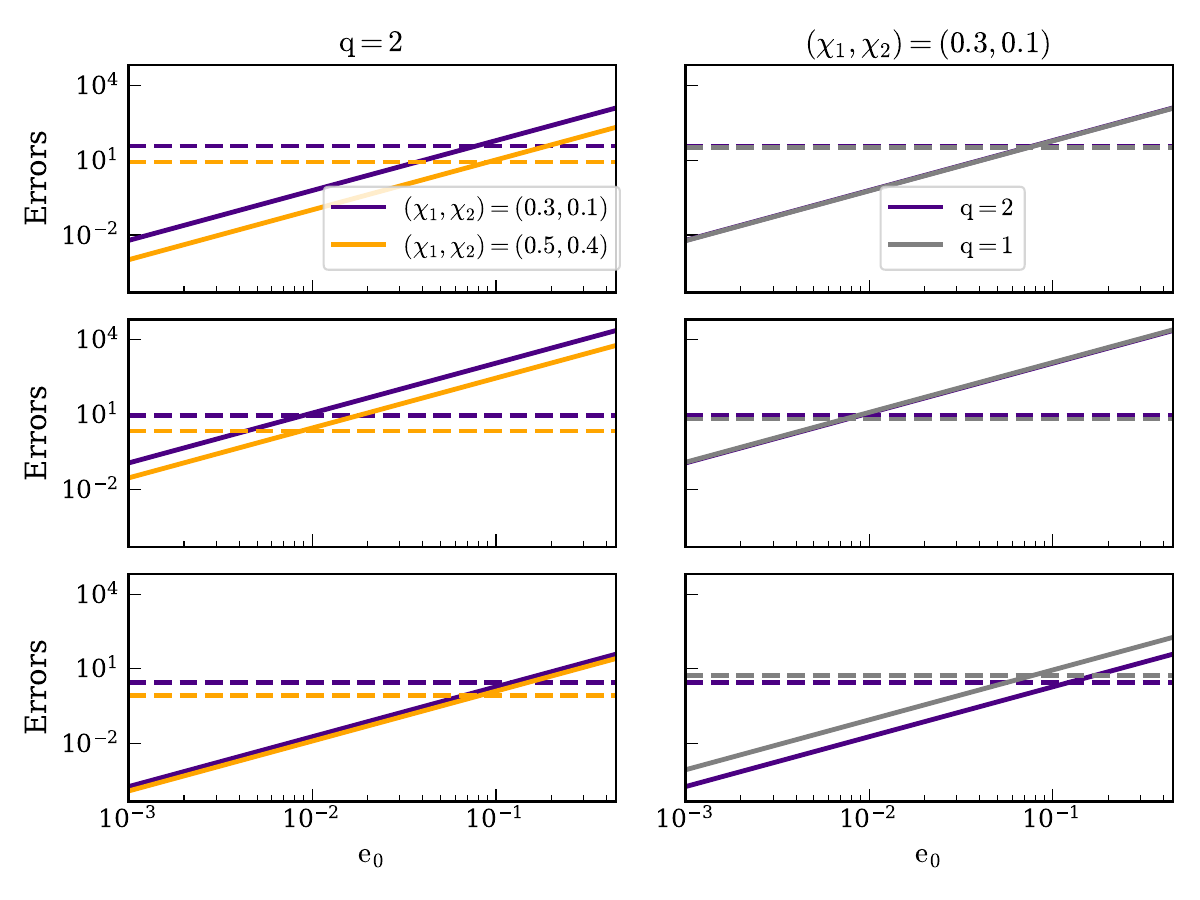}
    \caption{ The statistical error on the spin-induced quadrupole moment parameter $\sigma_{\delta\kappa_s}$ is plotted in dashed lines. The systematic error on the spin-induced quadrupole moment parameter $\Delta\theta_{\delta\kappa_s}$ is shown as a function of eccentricity. }
    \label{fig:syst_error}
\end{figure*}
%%%%%%%%%%%%%%%%%%%%%%%%%%%%%%%%%%%%%%%%%%%%%%%%%%%%%%%%%%

Figure~\ref{fig:syst_error} shows the systematic errors on the spin-induced quadrupole moment parameter, for LIGO-like sensitivity, Cosmic Explorer-like sensitivity and LISA sensitivity~\cite{KY19}. The dashed lines represent the statistical errors on the spin-induced quadrupole moment parameter $\sigma_{\delta\kappa_s}$. As expected, the systematic errors are lower than the statistical errors for smaller eccentricities in all cases. 

The first plot, representing the LIGO sensitivity, considers a compact binary system of masses $(10+5)M_{\odot}$ and dimensionless aligned spins $(\boldsymbol{\chi_1}, \boldsymbol{\chi_2})=(0.3, 0.1)$ and  $(\boldsymbol{\chi_1}, \boldsymbol{\chi_2})=(0.5, 0.4)$. As the figure shows, a slowly spinning system exhibits a larger statistical uncertainty in the spin-induced quadrupole moment measurement, consistent with previous findings~\cite{Krishnendu:2017shb}. However, the systematic errors overtake the statistical errors at similar eccentricity values, $e_0\sim0.1$. This means that for LIGO-like binaries, the eccentricity-induced biases are worrisome for binaries of initial eccentricity of $e_0\sim0.1$. In this case, the $e_0$ is defined as the eccentricity at $20 \rm Hz$.

When considering the third-generation detector configuration, such as Cosmic Explorer, as shown in the second plot, the eccentricity at which systematic errors become important shifts to a lower value, $e_0\sim0.01$, indicating the importance of including eccentricity effects in the upcoming detector era. For the Cosmic Explorer case, a binary of the same masses and spins is considered, fixing the $e_0$ value at $5\rm{Hz}$.

Finally, the eccentricity effects would be crucial for binaries in the LISA band at an eccentricity value, $e_0\sim 0.1$. For the LISA binary, though the spins are the same as those of previous cases, the masses are $(10^6, 10^5)M_{\odot}$. 

This study indicates the importance of incorporating orbital eccentricity effects while performing the spin-induced quadrupole moment-based BH nature tests. The Fisher-based systematic errors provide an order-of-magnitude estimate, and we will now explore the systematic bias in the measurements using a full Bayesian analysis.
\section{Systematic bias studies using Bayesian inference}
\label{sec:BayesianResults}

Following the Fisher information matrix-based systematic bias study, we will perform a full Bayesian analysis to quantify the biases introduced by orbital eccentricity in measuring the spin-induced quadrupole moment parameter of the binary.
\subsection{Bayesian inference and Parameter Estimation setting}
Bayes' theorem provides the posterior distributions on the parameters characterizing a compact binary system employing and is of the form, 
\begin{equation}
        p( \vec{\theta}\, | d,\, \mathcal{H}) = \frac{\pi( \vec{\theta}\, | \mathcal{H} ) \, \mathcal{L}(d\, |\, \vec{\theta}, \, \mathcal{H}) }{Z},
        \label{eq:posterior}
    \end{equation}
where $\vec{\theta}$ is the set of parameters, including the masses, spins, luminosity distance, and angles determining the location and orientations. In addition, the set of parameters in our case includes the spin-induced quadrupole moment parameters $(\delta\kappa_i, i=1,2)$ and the orbital eccentricity parameter. Equation~\ref{eq:posterior} provides posterior estimate on $\vec{\theta}$, given the data $d$ and hypothesis $\mathcal{H}$. The posterior probability distribution is obtained using the prior distribution $\pi( \vec{\theta}\, | \mathcal{H} )$, the likelihood function $\mathcal{L}(d\, |\, \vec{\theta})$ and the Bayesian evidence $Z$. 
The crucial ingredient to likelihood is the GW waveform model. The routine GW data inference assumes the detector noise to be Gaussian and stationary, leading to a Gaussian likelihood function. We can obtain the one-dimensional marginalized distribution on any parameter by marginalizing the likelihood function with all other parameters. For example, the one-dimensional marginalized posterior on the spin-induced quadrupole moment parameter is obtained as, 
\begin{equation}
        p(\delta\kappa\, \vert d,\, \mathcal{H} ) = \int p( \vec{\theta}\, | d,\, \mathcal{H})\, d\vec{\theta}.
        \label{eq:posterior-dks}
\end{equation}
The Bayesian evidence is a normalization constant and can provide the statistical significance of the hypothesis. 

We focus on a binary system of fixed total mass $15M_{\odot}$ and mass ratios $q=1, 3$. The dimensionless spin parameter is assumed to be oriented along the orbital angular momentum axis, with magnitudes  $(\boldsymbol{\chi_1}, \boldsymbol{\chi_2})=(0.3, 0.1)$ and  $(\boldsymbol{\chi_1}, \boldsymbol{\chi_2})=(0.5, 0.4)$. The luminosity distance to the source is fixed to $231 \rm{Mpc}$ such that the binary produces a signal-to-noise ratio of $70$ in the three-detector network, which consists of two advanced LIGO detectors along with the advanced Virgo detector~\cite{KAGRA:2013rdx, AdvancedLIGO2010, TheLIGOScientific:2014jea, TheVirgo:2014hva, TheVirgostatus} with advanced sensitivity~\cite{H1L1V1Dcc}. This choice ensures that the signal strength is sufficient to provide meaningful constraints on the spin-induced quadrupole moment parameters and the eccentricity. The binary BH simulations are produced using \texttt{TaylorF2Ecc} waveform model. Two different eccentricity values are considered to demonstrate the impact of orbital eccentricity: $e_0=0.1 \,(high)$ and $e_0=0.05\, (low)$, both defined at $20\rm Hz$.
\begin{table}[ht]
\centering
\begin{tabular}{ll}
\toprule
Run name & \hspace{4em}{Details} \\
\midrule
\texttt{run\_1} & $q=0.875$, $(\boldsymbol{\chi}_1, \boldsymbol{\chi}_2) = (0.3, 0.1)$ \\
\texttt{run\_2} & $q=0.5$, $(\boldsymbol{\chi}_1, \boldsymbol{\chi}_2) = (0.3, 0.1)$ \\
\texttt{run\_3} & $q=0.5$, $(\boldsymbol{\chi}_1, \boldsymbol{\chi}_2) = (0.5, 0.4)$ \\
\bottomrule
\end{tabular}
\caption{Summary of the simulations. The total mass is fixed to be $15M_{\odot}$. Two eccentricity values are considered: $e_0=0.1 \, (high)$ and $e_0=0.05\,(low)$.}
\label{tab:runs}
\end{table}
Table~\ref{tab:runs} contains the injection details and the specific notations used further in the paper. 

We consider two recovery models to estimate the potential bias in measuring spin-induced quadrupole moment parameters arising from neglecting orbital eccentricity. The first model incorporates eccentricity in the recovery. However, the second recovery model assumes quasi-circular orbits and disregards eccentricity. By comparing the parameters recovered from these two models, particularly the spin-induced quadrupole moment estimates, we evaluate the impact of eccentricity on parameter inference. We also study how including or excluding eccentricity affects the recovery of intrinsic binary parameters such as the chirp mass, $\rm M_c$, by performing binary BH parameter estimation for both circular and eccentric orbits. In all cases, the spin-induced quadrupole moment parameters are varied simultaneously during the recovery. That is, $\delta\kappa_1$ and $\delta\kappa_2$ are treated as free parameters alongside the standard binary BH parameters.
\subsection{Tests of BH nature: Eccentric versus circular orbits}
%%%%%%%%%%%%%%%%%%%%%%%%% FIGURE
\begin{figure}
    \includegraphics[width=3.5 in]{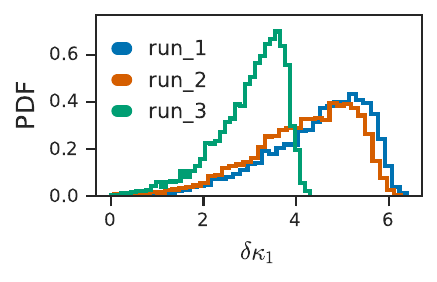}
    \caption{For three binary choices, $q=0.875$, $(\boldsymbol{\chi}_1, \boldsymbol{\chi}_2) = (0.3, 0.1)$ (blue), $q=0.5$, $(\boldsymbol{\chi}_1, \boldsymbol{\chi}_2) = (0.3, 0.1)$ (orange), $(\boldsymbol{\chi}_1, \boldsymbol{\chi}_2) = (0.3, 0.1)$ and $q=0.5$, $(\boldsymbol{\chi}_1, \boldsymbol{\chi}_2) = (0.5, 0.4)$ (green), $\delta\kappa_1$ is estimated considering a simulated binary system in eccentric orbit with initial orbital eccentricity $e_0^{\rm inj}=0.1$ and recovered employing a quasi-circular waveform model. The mass ratio effect is negligible. However, a slowly spinning binary exhibits a larger bias compared to a rapidly spinning binary.}
    \label{fig:kappa_1_ecc_bias}
\end{figure}
%%%%%%%%%%%%%%%%%%%%%%%%% FIGURE
\begin{figure}
    \includegraphics[width= 3.5 in]{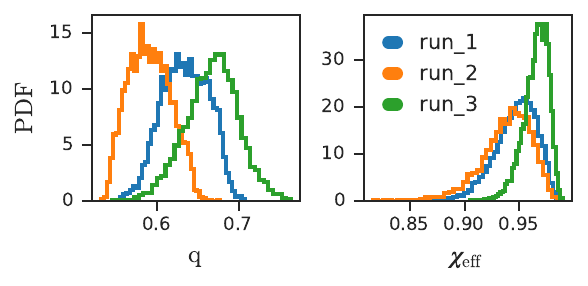}
    \caption{The posteriors on the mass ratio and effective spin parameter are shown for three binary configurations. The injected signals are eccentric, but are analysed using a quasi-circular waveform model that includes the $\delta\kappa$ parameter. The injected mass ratio values are $q = 0.875,0.5,0.5$, and the corresponding effective spin values are $ \boldsymbol{\chi_{\rm eff}} = 0.20, 0.23, 0.46 $, plotted in green, orange, and blue curves, respectively. The posterior estimates are shifted from their injected values, with the extent of the bias determined by the respective binary configuration.}
    \label{fig:mass_ratio_chi_eff_posteriors}
\end{figure}
%%%%%%%%%%%%%%%%%%%%%%%%% 
%%%%%%%%%%%%%%%%%%%%%%%%% FIGURE
\begin{figure*}
    \includegraphics[width=\textwidth]{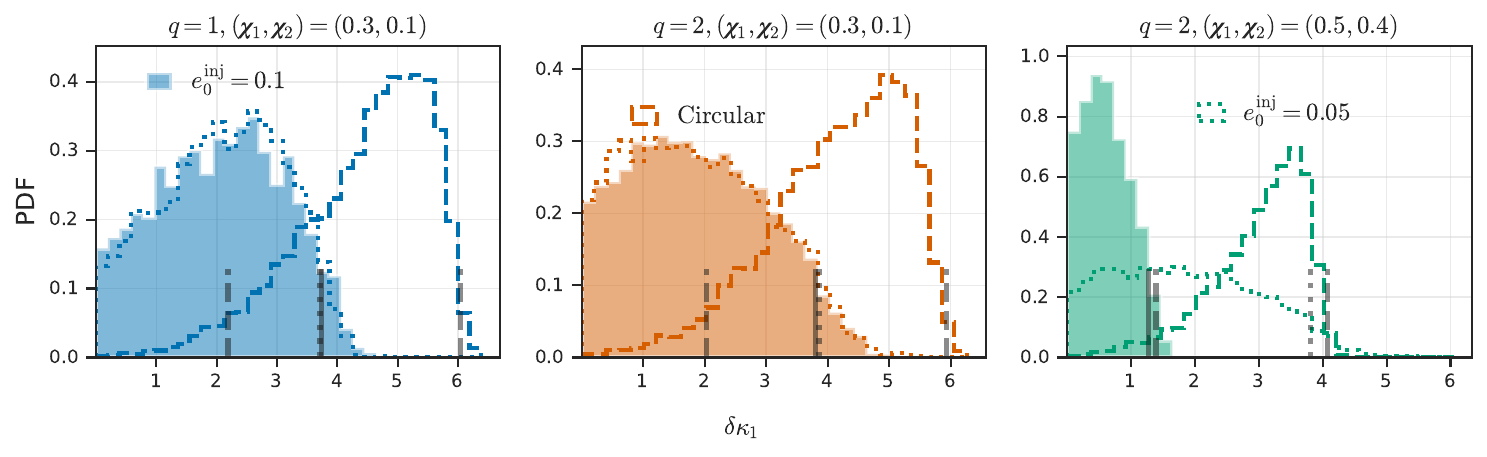}
    \caption{Posteriors on $\delta\kappa_1$ considering a simulated binary of eccentricity $e_0^{\rm inj}=0.1$ (solid histogram) and $e_0^{\rm inj}=0.05$ (dotted histograms). Both these injections, when recovered with the eccentric model, provide results consistent with a binary BH value. In contrast, the circular recovery of the $e_0^{\rm inj}=0.1$ injection lead to severe bias in the $\delta\kappa_1$ estimate (dashed histogram). The vertical lines represent the $90\%$ credible highest probability density regions.}
    \label{fig:kappa_1_ecc_cir_comp}
\end{figure*}
%%%%%%%%%%%%%%%%%%%%%%%%% FIGURE
%\begin{figure*}
%    \includegraphics[width=\textwidth]{kappa_1_2_cir_comp.png}
%    \caption{Posteriors on $\delta\kappa_1$ and $\delta\kappa_2$ are compared when a circular recovery is performed on a eccentric binary injection. Though the $\delta\kappa_1$ parameter shows major biases, the $\delta\kappa_2$ posteriors are well constrained to the BH value.}
%    \label{fig:kappa_1_2_cir_comp}
%\end{figure*}
{\it Systematic shift in $\delta\kappa_1$ estimates:} 
The standard spin-induced quadrupole moment tests of BH nature assume that binaries are in quasi-circular orbits. In such cases, posterior distributions peaking at zero are interpreted as consistent with the BH hypothesis, while significant deviations may suggest evidence for non-BH compact objects. However, we find that the effects of orbital eccentricity can produce similar shifts in the posteriors. This implies that an eccentric binary can mimic the signatures expected from a non-BH signature when analysed using a quasi-circular waveform model. As a result, eccentricity can lead to false positives in tests of the BH nature.

Figure~\ref{fig:kappa_1_ecc_bias} shows the posterior probability distributions on $\delta\kappa_1$ for three cases:  nearly equal mass slowly spinning binary (blue), asymmetric slowly spinning binary (orange) and nearly equal mass moderately spinning binary (green). The injections assume an eccentric binary with $e_0^{\rm inj}=0.1$ and a total mass of $15M_{\odot}$. The recoveries employ a quasi-circular waveform model. The mass ratio appears to have a minor impact on the bias in the $\delta\kappa_1$ estimate arising from neglecting orbital eccentricity. In contrast, spin magnitudes play a major role. We find that systems with lower spin magnitudes exhibit larger biases in the $\delta\kappa_1$ posterior, whereas higher-spin configurations yield reduced bias and tighter posteriors. This highlights the importance of spin in accurately constraining spin-induced quadrupole moments in the presence of orbital eccentricity.

{\it Impact on individual binary parameters:} 
Neglecting orbital eccentricity in performing tests of BH nature not only impacts the $\delta\kappa_1$ but also the intrinsic binary parameters, specifically, let us evaluate the impact on mass ratio,
\begin{equation}
    q=m_2/m_1, m_1>m_2
\end{equation}
and the effective spin parameter, 
\begin{equation}
    \boldsymbol{\chi_{\rm eff}} = \frac{m_1\boldsymbol{\chi_1}+m_2\boldsymbol{\chi_2}}{m_1+m_2}.
\end{equation} 
The three binary configurations considered have true values of mass ratio $q=0.875, 0.5, 0.5$ and effective spin $\boldsymbol{\chi_{\rm eff}}=0.20,0.23,0.46$, as inferred from the individual component masses and spin magnitudes. 
As shown in Fig.~\ref{fig:mass_ratio_chi_eff_posteriors}, the posterior estimates are significantly biased relative to their injected values. The mass ratio estimates are $q=0.635^{+0.046}_{-0.043}, 0.590^{+0.041}_{-0.041}, 0.669^{+0.050}_{-0.057}$ and effective spin estimates are $\boldsymbol{\chi_{\rm eff}}=0.949^{+0.028}_{-0.031}, 0.942^{+0.032}_{-0.037}, 0.967^{+0.017}_{-0.018}$ at $90\%$ credible interval. The uncertainties are calculated relative to the posterior medians.  This indicates that the spin magnitude effect is more dominant than the mass ratio effect. Also, this mischaracterisation leads to overly constrained posteriors on $\boldsymbol{\chi_{\rm eff}}$ which, in turn, results in correspondingly stringent estimates of the spin-induced quadrupole parameter $\kappa_1$, suggesting the importance of including orbital eccentricity in tests of BH nature and inference of their astrophysical properties.

{\it Comparing the impact of $e_0^{\rm inj}=0.1$ and $e_0^{\rm inj}=0.05$:} 
As demonstrated in Sec .~\ref {sec:systematic_errors}, the estimates are unaffected if the eccentricity is small enough. Figure~\ref{fig:kappa_1_ecc_cir_comp} compares the $\delta\kappa_1$ estimates from circular recovery (dashed curves), eccentric recovery for $e_0^{\rm inj}=0.1$ (solid curves) and eccentric recovery for $e_0^{\rm inj}=0.05$ (dotted curves). Both the eccentric recoveries provide unbiased estimates, as expected. The $90\%$ CI, highest posterior density (HPD) bounds are marked in black vertical lines. 

{\it Effect on mass ratio and effective spin parameter:} Figure~\ref{fig:q_chieff_scatter} shows the scatter plot between the $\boldsymbol{\chi}_{\rm eff}$ and $q$, for an eccentric binary injection with $e_{0}^{\rm inj}=0.1$. The left plot does not include spin-induced quadrupole moment parameters, and the effect of neglecting eccentricity is demonstrated on a binary BH injection and recovery model. The red and blue scatters represent eccentric recovery and circular recovery, respectively. Whilst the right figures consider $\delta\kappa$ in the recovery model, the injections are the same in both cases. The red scatter in both cases includes the injected value, as indicated by the $star$ symbol. Notably, excluding eccentricity in the spin-induced quadrupole moment analysis can lead to serious biases if the true signal is in an eccentric orbit. For $e_{0}^{\rm inj}=0.1$, the nearly equal mass binary is wrongly estimated as a mass-asymmetric binary with rapid spins. As a consequence, the estimation of binary astrophysical population properties will also be biased. 
%%%%%%%%%%%%%%%%%%%%%%%%% FIGURE
\begin{figure*}
    \includegraphics[width=\textwidth]{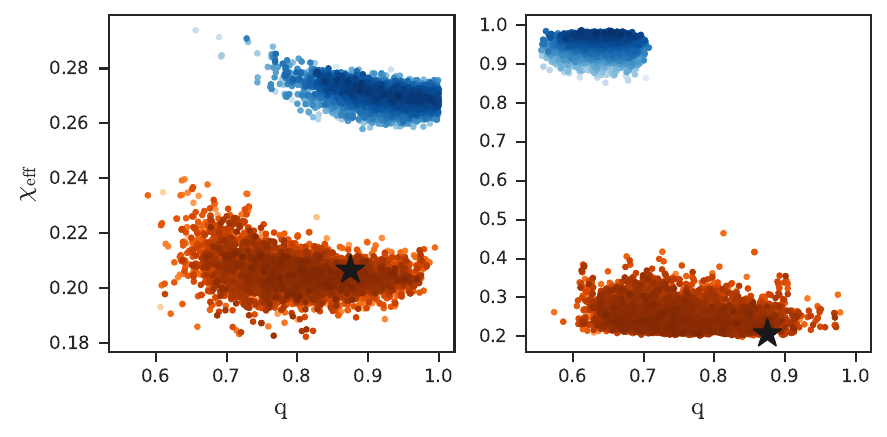}
    \caption{Impact of excluding eccentricity while analysing an eccentric binary system is illustrated for two cases, for a binary BH analysis (left, $\theta_{\rm BBH}$) and SIQM analysis (right, $\theta_{\rm BBH}, \delta\kappa_i$). The black star represents the injected value. Red and blue scatter plots represent analysis assuming circular and eccentric recovery, respectively. Ignoring eccentricity in the SIQM analysis can lead to mis-identifying a nearly-equal mass, slowly spinning binary BH as a mass-asymmetric binary with evidence for rapid spins.}
    \label{fig:q_chieff_scatter}
\end{figure*}
\subsection{Implications to compact binary formation and evolution}
Accurate inference of the binary mass, spin, and eccentricity parameters is essential to fully understanding the astrophysical properties and correctly interpreting the fundamental physics. Waveform models with missing physics, such as orbital eccentricity, can lead to incorrect estimates of binary intrinsic and extrinsic parameters. In addition, employing such models can lead to incorrect interpretation of tests of the BH nature. In our case, we find a clear indication of a non-BH compact object when analysing a binary BH signal in an eccentric orbit using a quasi-circular template. Therefore, ignoring eccentricity can result in a completely different astrophysical formation scenario. An isolated binary evolution can be misinterpreted as a dynamical formation when the original equal-mass, moderately spinning binary is incorrectly identified as a rapidly spinning, asymmetric binary due to modelling errors. In addition, notice that the formation and evolution of non-BH compact objects is still need to be understood. 
\section{Discussion and Summary}
\label{sec:summary}
The spin-induced quadrupole moment–based test of BH nature is a routinely employed method to unravel the true nature of detected compact objects. While the measurability of spin-induced quadrupole moments for binaries in quasi-circular orbits has been extensively studied, there has been no systematic study for binaries in eccentric orbits. In this work, we investigate the impact of neglecting orbital eccentricity in performing tests of BH nature, focusing on the spin-induced quadrupole moment test for the first time. We find that neglecting eccentricity can lead to significant biases in determining the true nature of compact objects. Notably, a binary with total mass $15 M_{\odot}$, producing an SNR of about $70$ and orbiting with an initial eccentricity of $0.1$, can lead to exclusion of the BH value for the $\delta\kappa_1$ parameter at $90\%$ HPD. When analysed using a quasi-circular template, this system yields a biased estimate of $\delta\kappa_1 = 6^{+0.3}_{-0.3}$ at the 90\% HPD credible interval. This highlights the possibility of misidentifying a binary BH in an eccentric orbit as a binary boson star in a circular orbit. In addition, the binary mass and effective spin parameters will also be incorrectly measured in this case. For example, instead of recovering the true values of $(8+7) M_{\odot}$ and $(\boldsymbol{\chi}_1, \boldsymbol{\chi}_2) = (0.3, 0.1)$, the quasi-circular model recovers biased values of $(9.7+6.1) M{\odot}$ and $(\boldsymbol{\chi}_1, \boldsymbol{\chi}_2) = (0.96, 0.93)$, when analyzing a eccentric injection with $\delta\kappa$ parameters in the recovery.

This is the first study to quantify eccentricity-induced systematic biases and detail the astrophysical implications, considering generic binary BH signals. The method developed is also applicable to binaries consisting of neutron stars.
However, in the case of distinguishing neutron stars from non-BH objects, one must take into account the tidal deformability effect. In most cases, the neutron star spin can be neglected. Eccentricity plays a crucial role in neutron star–BH binaries. For such systems, the higher harmonics in the waveform model are also important due to the large mass asymmetry. This study demonstrates the systematic errors by devising a semi-analytic Fisher matrix-based approach and further establishes the findings using full Bayesian inference. The methods are generic and can be employed for future observations and waveform models.

The waveform model used describes an inspiralling compact binary system with the spin angular momenta either aligned or anti-aligned with the orbital angular momentum of the system. In the future, as an extension of the current method, a model including spin-induced orbital precession effects with orbital eccentricity effects, such as the ones used in~\cite{Morras:2025nlp}, will be employed. The {\texttt{TaylorF2Ecc}} model treats the spin dynamics and the eccentricity effects separately. However, the method can also be extended by using models that calculate the eccentricity effects for spinning BHs~\cite{Sridhar:2024zms} and will be explored in a future study. These more accurate models will not change the paper's findings. However, they can provide precise estimates of the eccentricity-induced biases in the spin-induced quadrupole moment-based test of the nature of BHs.
\acknowledgements 
Krishnendu is supported by STFC grant ST/Y00423X/1. The computations described in this paper were performed using the University of Birmingham's BlueBEAR HPC service, which provides a High Performance Computing service to the University's research community. See http://www.birmingham.ac.uk/bear for more details. Krishnendu thanks Khunsang Phukon for carefully reading the manuscript and providing useful comments and suggestions. This document has LIGO preprint number {\tt P2500663}. 
\bibliographystyle{apsrev}
\bibliography{ref}
\end{document}